%% file: main.tex
\documentclass[conference]{IEEEtran}
\input{format}

\newcommand\copyrighttext{%
  \footnotesize \textcopyright 2022 IEEE. Personal use of this material is permitted.
  Permission from IEEE must be obtained for all other uses, in any current or future 
  media, including reprinting/republishing this material for advertising or promotional 
  purposes, creating new collective works, for resale or redistribution to servers or 
  lists, or reuse of any copyrighted component of this work in other works.
  }
\newcommand\copyrightnotice{%
\begin{tikzpicture}[remember picture,overlay]
\node[anchor=south,yshift=10pt] at (current page.south) {\fbox{\parbox{\dimexpr\textwidth-\fboxsep-\fboxrule\relax}{\copyrighttext}}};
\end{tikzpicture}%
}

\begin{document}

\title{RESAM: Requirements Elicitation and Specification for Deep-Learning Anomaly Models with Applications to UAV Flight Controllers}

\author{\IEEEauthorblockN{Md Nafee Al Islam, Yihong Ma, Pedro Alarcon, Nitesh Chawla, Jane Cleland-Huang}
 \IEEEauthorblockA{Dept. of Computer Science and Engineering} 
 University of Notre Dame\\
 Notre Dame, IN, USA\\
 mislam2@nd.edu, JaneHuang@nd.edu}

\maketitle

\begin{abstract}
CyberPhysical systems (CPS) must be closely monitored to identify and potentially mitigate emergent problems that arise during their routine operations. However, the multivariate time-series data which they typically produce can be complex to understand and analyze. While formal product documentation often provides example data plots with diagnostic suggestions, the sheer diversity of attributes, critical thresholds, and data interactions can be overwhelming to non-experts who subsequently  seek help from discussion forums to interpret their data logs.  Deep learning models, such as Long Short-term memory (LSTM) networks can be used to automate these tasks and to provide clear explanations of diverse anomalies detected in real-time multivariate data-streams. In this paper we present RESAM, a requirements process that integrates knowledge from domain experts, discussion forums, and formal product documentation, to discover and specify requirements {and design definitions in the form of time-series attributes that contribute to the construction of} effective deep learning anomaly detectors. We present a case-study based on a flight control system for small Uncrewed Aerial Systems and demonstrate that its use guides the construction of effective anomaly detection models whilst also providing underlying support for explainability. {RESAM is relevant to domains in which open or closed online forums provide discussion support for log analysis.}
\end{abstract}
\copyrightnotice

\input{sec_Introduction}

\input{sec_process}
\input{sec_case_study}

\input{sec_common_failures}
\input{sec_techniques}

\input{sec_evaluation}
\input{sec_discussion}
\input{sec-related}

\section{Conclusions and Future Work}
\label{sec:conclusions}
In this paper, we have presented \RM as a systematic approach for discovering, specifying, and analyzing requirements for DLMs. Diagnostic information provided by formal documentation was augmented by expert opinions shared in online discussion forums. We combined the information from both sources to discover and specify requirements for anomalies related to vibration, mechanical problems, compass interference, GPS glitches, and power problems. We then leveraged this information to design and train  LSTM time-series models for three failure types and compared the results to rule-based anomaly detectors. Our evaluation showed that the requirements specified as a result of the \RM were sufficient for guiding the construction of effective anomaly detectors. Furthermore, we observed that the expert knowledge obtained from the forum discussions led to anomaly detectors that outperformed the ones based on recommendations from the formal documentation alone.
Finally, we have laid out a series of open challenges including automated support for mining discussion forums, exploring more diverse algorithms, providing explanations for the identified anomalies, and specifying and analyzing non-functional requirements to guide the architectural decisions of the anomaly detectors.

 \section*{Acknowledgments}
This work is primarily funded by the USA National Aeronautics and Space Administration (NASA) under grant number: 80NSSC21M0185.

\bibliographystyle{IEEEtran}
\interlinepenalty=10000
\bibliography{IPSN} 

\end{document}

%% file: format.tex
\usepackage{url}
\usepackage{multirow}
\usepackage{array} 
\usepackage{amsmath}
\usepackage{amssymb}
\usepackage{graphicx}
\usepackage{flushend}
\usepackage{enumitem}
\usepackage{xcolor}
\usepackage{svg}
\usepackage{svg-extract}
\usepackage{subcaption}
\usepackage{xspace}
\usepackage{tabulary}
\usepackage{booktabs}
\usepackage{listings}
\usepackage{tabularx}
\usepackage{colortbl}
\usepackage[export]{adjustbox}
\usepackage{comment}
\usepackage[ruled,vlined]{algorithm2e}
\usepackage{algpseudocode}
\usepackage{booktabs}
\usepackage{multicol,multirow}
\usepackage{lscape}
\usepackage{fdsymbol}
\usepackage{wrapfig}
\usepackage{arydshln}
\usepackage{soul}

\usepackage{tikz}
\usepackage{textcomp}
\usepackage{lipsum}

\definecolor{grn}{HTML}{00FF00}
\definecolor{ylw}{HTML}{FFFF00}
\definecolor{pblue}{HTML}{AFCFEE}

\definecolor{pink}{rgb}{0.9,0,0.9}
\definecolor{gray}{rgb}{0.95,0.95,0.85}
\definecolor{mauve}{rgb}{0.1,0.7,0.2}
\definecolor{dkgreen}{rgb}{0,0.6,0}
\definecolor{lightgray}{rgb}{0.6,0.6,0.6}

\newcommand{\RM}{\emph{RESAM~}}
\newcommand{\RMx}{\emph{RESAM}}

\newcommand*{\mybox}[2]{\colorbox{#1!30}{\parbox{.98\linewidth}{#2}}}
\newcolumntype{M}[1]{>{\centering\arraybackslash}m{#1}}
\newcolumntype{L}[1]{>{\raggedright\let\newline\\\arraybackslash}p{#1}} 
\newcolumntype{C}[1]{>{\centering\let\newline\\\arraybackslash}p{#1}} 
\usepackage{fdsymbol}
\usepackage{booktabs} 
\usepackage{color}
\usepackage{wasysym}
\usepackage{hyperref}

\usepackage{fdsymbol}

%% file: sec_introduction.tex
\section{Introduction}
\label{sec:Introduction}
Cyber-Physical Systems (CPS), supported by IoT devices, are becoming increasingly ubiquitous in society where they are used for diverse sensing and actuating tasks such as detecting environmental conditions, supervising power plants and factory environments, and monitoring autonomous machines such as factory floor robots and small Uncrewed Aerial Systems (sUAS). These IoT devices often collect large  amounts of multi-variate time-series data which are used in realtime or offline to detect anomalies and to diagnose failure conditions. Analyzing this data often requires significant domain expertise. Some market products include tools that are designed to help users interpret the data. For example, in the commercial sector, AssetSense collects, monitors, and analyzes data from power plants and provides interactive tools to allow users such as operators, managers, and other plant personnel, to effectively interpret the data and to make actionable decisions \cite{assets}. However, in other domains, huge volumes of data are collected without the benefit of fully automated analysis tools.

The use of deep learning models (DLMs) to monitor data streams and create user-facing explanations of anomalies can address this need.  However, while DLMs can learn to recognize anomalies through a self-supervised training process, they often have a limited ability to explain the impact and root causes of the anomalies that they detect -- especially for less common cases or cases in which multiple errors occur simultaneously and explanations are quite complex. On the other hand, domain experts often have deep knowledge about the attributes and multivariate time-sequence patterns that serve as indicators of those failures. By leveraging this knowledge we can design and deploy ensembles of effective DLMs, each targeted at different types of anomalies and imbued with the ability to explain the anomalies they detect. However, this introduces a requirements engineering (RE) challenge to determine what types of anomalies are worth detecting and explaining, and identifying the right data sources for training DLMs to detect them \cite{REChallengeInML}. In practice, requirements for DLMs often go under-explored, and as a result the training data is often not well-aligned with the goals of the DLM. Therefore, the research challenge that we address in this paper focuses on {\bf``how to systematically leverage available information sources in order to identify, analyze, and specify {functional} requirements for designing Deep Learning Models capable of detecting and explaining anomalies in multivariate time-series data?''} This represents an emergent challenge at the intersection of requirements engineering and AI, and is likely to increase in importance with the growing use of Machine Learning (ML) and AI solutions to support the rapidly expanding deployment of IoT devices  \cite{DBLP:conf/re/VogelsangB19}. {While other aspects of the requirements elicitation process, such as  non-functional requirements (NFRs) related to critical qualities of fairness, accuracy, and performance, are equally important, this paper focuses primarily on the functional aspects.}

{Our approach can potentially be useful in any domain for which an active discussion forum provides support for analyzing data streams or logs. Examples include the sUAS domain \cite{qground,mplanner,ardupilot}, automotive data logs \cite{autodomain}, and space observatory data (e.g., \cite{astronomydomain})}. However, rather than address this challenge in a general way, we take a relatively deep dive into the domain of sUAS flight controllers, motivated by the challenges we have experienced in designing and constructing DLMs for sUAS flight control data, coupled with the lack of a prescriptive requirements process describing how to accomplish this task. 

\begin{figure*}[t]
    \centering

    \includegraphics[width=.98\textwidth]{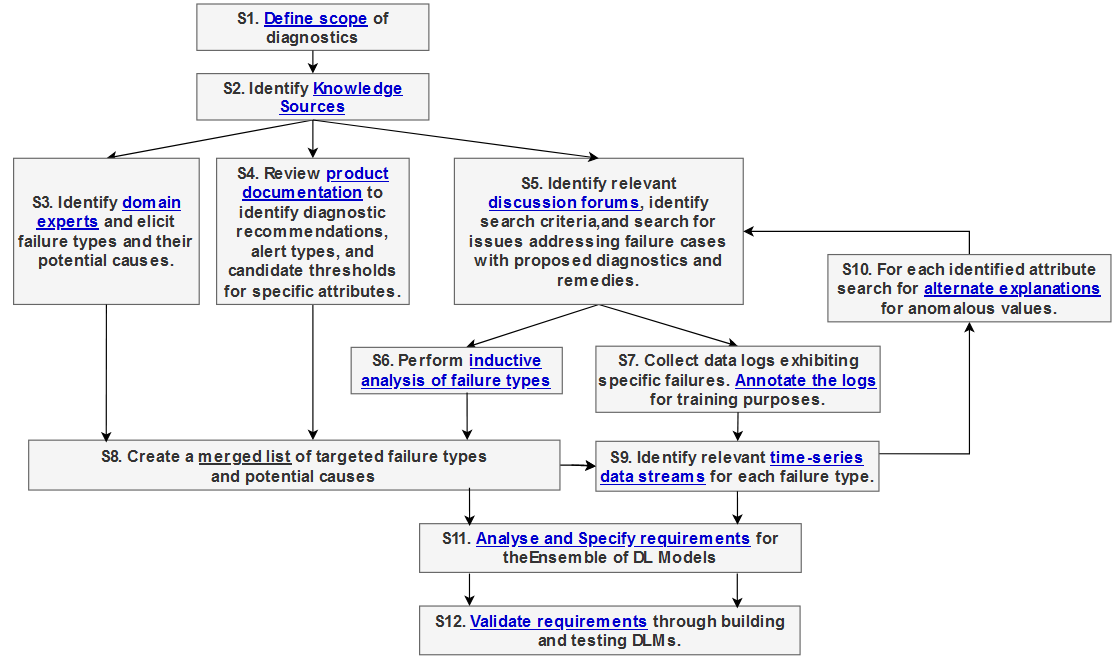}
    \caption{RESAM process for supporting the design and validation of deep-learning anomaly detection models }
    \label{fig:requirements_model}
\end{figure*}

Our paper therefore makes two primary contributions. First, it describes \RMx, a systematic requirements process for leveraging multiple data sources including domain experts, discussion forums, and documentation, in order to elicit, analyse, and specify requirements for DLM-based monitoring and analysis tools. \RM stands for {\bf R}equirements {\bf E}licitation and {\bf S}pecification for {\bf A}nomaly {\bf M}odels.  Second, given the domain of sUAS, we provide example specifications for constructing anomaly detectors for five different types of failures. Then, to validate \RMx, we use the specified requirements to guide the construction of deep-learning LSTM (Long short-term memory) models and show that these models satisfy their specified requirements.  Further, given the effort of mining requirements information from discussion forums, we compare the quality of the DLMs built using forum data versus those built using only formal documentation, and show that in our experiments incorporating forum data results in more accurate models.

The remainder of this paper is structured as follows. Section \ref{sec:process} describes our \RM approach for identifying failures, mapping them to sensor attributes, and specifying initial requirements for DL-based anomaly models. Section \ref{sec:uavstudy} adopts the \RM process to specify requirements for five types of anomaly. Sections \ref{sec:techniques} and \ref{sec:eval} validate the \RM requirements through constructing and testing two types of models. Finally, Sections ref{sec:discussion} to \ref{sec:conclusions}  discuss the strengths and limitations of our work, identify open challenges, discuss related work, and draw conclusions.

%% file: sec_process.tex
\section{The RESAM Process}
\label{sec:process}
Given the complexity of sensor data produced by many CPS and IOT devices, it is helpful to follow a systematic requirements process for discovering, specifying, analyzing, and validating requirements for DLMs that are capable of effectively detecting and explaining anomalous behavior.  Models built in a more ad-hoc manner can suffer from several different problems including: (1) incomplete analysis of failure types that the DLM should be trained to detect, (2) incomplete analysis of the root causes of a specific failure type resulting in missed attributes and an unreliable model that fails to detect key anomalies, and finally (3) n\"aive diagnoses that fail to consider alternate explanations for anomalous values and therefore return incorrect diagnoses.

We lay out our systematic process through steps (S1-S12) in Figure \ref{fig:requirements_model}. First we establish initial scope on what the system will and will not do (S1) {through the requirements activities of \emph{stakeholder identification}, \emph{and preliminary requirements elicitation} in order to \emph{establish goals} and \emph{system boundaries} for the planned AI components} \cite{robertson}. Leveraging the knowledge of the stakeholders, we then identify knowledge sources for supporting the  requirements discovery process {associated with each of the planned AI components }(S2). These sources include \emph{domain experts} who brainstorm a list of common failures and potential causes (S3), \emph{formal product documentation} including any information about diagnostic checks and alerts (S4), and relevant \emph{discussion forums} (S5). Discussion forums can be public or private. Either way they can contain potentially vast quantities of data typically grouped by categories and threads. We therefore establish search criteria designed to identify specific \emph{types of failures}, and perform inductive analysis to identify failure types (S6). Depending upon the nature of the forum, some posts will include data logs with accompanying analysis (S7). These logs are helpful for several reasons including (a) identifying failure types, (b) discovering how to detect failures of each type, and (c) building a collection of failure examples for later training of the DL models. We therefore annotate these logs according to expert consensus achieved in the forum discussion.

Based on the results from each of these steps we create an initial, merged list of failure types (S8), triage the failures to establish the scope of the DL models and then identify relevant time-series data streams for each failure type (S9). These steps represent a top-down approach to requirements discovery which can easily miss alternate explanations for a specific anomaly. For example, we might find that a specific type of failure is common and has symptoms A and B, but we could miss the fact that A is also a symptom of other types of failures. Our approach therefore also includes a bottom-up search for forum posts that discuss each identified symptom (S10) in order to explore alternate explanations for the symptoms. 
Once failures, attributes, and patterns have been identified, we specify  and analyze requirements for the targeted anomaly models (S11), and finally train, validate, and test the DL models using the previously annotated datasets to ensure that they satisfy the requirements (S12). 

%% file: sec_case_study.tex
\section{Applying RESAM to an sUAS Flight Controller}
\label{sec:uavstudy}

We illustrate and evaluate our approach in the domain of sUAS flight controllers \cite{sensing, surveillance1}, where the growing number of sUAS has led to a rapid escalation of incidents \cite{accident1, accident2}.  {Our targeted stakeholders primarily include remote pilots in command (RPICs), sUAS maintainers, and accident investigators.} Many flight incidents have occurred due to faults caused by GPS glitches, incorrect configurations, or mechanical problems. When problems are experienced, remote sUAS pilots use software applications, such as QGroundControl \cite{qground} and MissionPlanner \cite{mplanner} to inspect the flight logs in an attempt to diagnose the problem. However, given thousands of configuration parameters and sensor readings in typical flight controllers, these tools tend to focus on common failure cases whilst lacking specific information about data attributes, acceptable threshold values, and multivariate patterns that are symptoms of emergent problems. As a result, remote pilots (many of them hobbyists) often struggle to interpret flight log data, and often resort to posting their flight logs on open-source forums in order to get advice from experts. As experts are often able to quickly pinpoint the root cause of the failure, our goal is to leverage their advice from the discussion forums, to specify, design and construct DLM anomaly models.  We focus on the \emph{Ardupilot} \cite{ardupilot} flight controller, as it represents one of the most popular open-architecture flight control stacks.

\subsection{Analyzing Formal Documentation (S4)}
The Ardupilot documentation provides a clear list of common failures found under two categories of {\it Problem Diagnosis}\footnote{
\url{https://ardupilot.org/copter/docs/common-diagnosing-problems-using-logs.html}} and {\it Prearming safety checks}\footnote{\url{https://ardupilot.org/copter/docs/common-prearm-safety-checks.html}}. We list these in Table \ref{tab:fault_types} as common {diagnosis} problems (CP) and safety checks (SC) respectively. The most common diagnosis problem include  excessive vibration,  mechanical failure,  power issues,  GPS failures,  compass interference, and  failsafe errors.  The ArduPilot documentation on safety checks (SC) describes several problems also covered in the Problem Diagnosis explanations. In addition it describes  Barometer failures,  Compass failures, and Inertial Navigation System failures. For each of these failure types, we retrieved lists of attributes and ranges of their values that served as indicators or symptoms of the failure.

The Ardupilot documentation also provides mappings between some of the failure types and specific attributes that serve as {\it indicators} of failure.
 It documents 19 different groupings of attributes, such as 
ATT (Attitude), CTUN (Control, Throttle, and altitude), and 
IMU (Accelerometer and gyro information). Each group contains from 1 to 20 attributes. For example, ATT includes actual and desired values for Roll, Pitch, and Yaw; while CTUN includes ThO (throttle). While the documentation provides mappings from failures to specific attributes, details are often sparse.  For example, in cases such as {\it Gyros not healthy}, the problem is described as ``One of the gyroscopes is reporting it is unhealthy which is likely a hardware issue''; however no attributes are listed as indicators of poor gyro health. In many cases the documentation failsto describe normal thresholds or ranges, most likely because these are expected to vary across different models of sUAS. As a stand-alone knowledge source, it therefore lacks sufficient details for guiding the development of a broad set of DLM for purposes of anomaly detection. 

\input{tables/faults.tex}
Given insufficient documentation, RPICs are often confused about the underlying root cause of a problem.  For example, one person posted a comment to the Ardupilot forum (with expletives removed) saying that ``I am afraid to arm because all kinds of warnings pop up: at first it was HIGH GPS HDOP..., then ALT DISPARITY... : BAD GYRO HEALTH.... {\bf I have no idea what to do.}'' Forum experts were able to discuss and diagnose this problem as a power supply issue, meaning that none of the actual error messages were directly related to the diagnosis. In general we found that the Ardupilot documentation provided information about the most common causes of failures, whilst ignoring edge cases which occur frequently enough to cause problems in practice. 

\subsection{Analyzing Forum Discussions (S5)}
Given these limitations, we conducted a systematic search of issues in Ardupilot discussion forums\footnote{\url{https://discuss.ardupilot.org}} with the aim of augmenting the information provided in the formal documentation in order to identify and specify clear requirements for building our targeted suite of anomaly detectors.  We first selected the `Log Analysis' category and then used search terms of{ vibration}, { power}, { motor}, { mechanical}, { compass}, { interference}, { GPS}, {  battery}, and { roll}, based on the emphasis on these problems in the formal documentation.  We then filtered out the results to remove issues that (1) represented basic queries about accessing or viewing logs such as ``How are the logs stored?'' or ``how to set the camera to record crashes?'', (2) provided only vague descriptions of the problem (e.g., ``drone is acting weird''), (3) had neither logs attached nor graphs describing the problem, or (4) lacked a clear diagnosis of the problem in the subsequent discussion. 
For each of the retrieved issues, we carefully read the post and the subsequent discussion, and reviewed data plots using the Mission Planner and MAV Explorer tools. While there is a possibility that some of the  recommendations were incorrect, they were frequently corroborated by others. Approximately 25\% of the top ranked solutions were posted by users with a forum-wide average of 304 likes, and were owners of with `developer' badges with an average of 83.5 posts each.

There were numerous cases in which experts provided insightful comments. For example, in response to a help request for interpreting an in flight EKF-INAV failsafe warning\footnote{{https://discuss.ardupilot.org/t/beginning-flight-testing-log-analysis-shows-in-flight-failsafe-ekfinav/64970}}, one expert identified several different issues including vibration on the Z axes, a problem with the motor PWM, and a compass interference problem. Such insights enhanced the knowledge provided by the formal documentation, especially when experts suggested alternate causes to those already described in the documentation. 

To identify an initial list of failures, we classified each issue we encountered using a hybrid deductive/inductive approach. If the issue referred to one of the failures described in the Autopilot documentation (cf. Table \ref{tab:fault_types} CP and SC), then we categorized it accordingly. All other reported issues that did not match any existing category were initially classified as { `other'}. For issues in the { `other'} category we used a snowballing technique which used key terms from the discussion to search for other related posts. For example given a post describing { failure to disarm} we searched for {`problem'} and {`autodisarm'}. This resulted in 36 posts categorized as {`other'}. We manually clustered  similar issues and then named the  cluster appropriately. This produced the following six additional types of failures: {`tuning errors'}, {`EKF switching'}, {`voltage sensor calibration'}, {`logging issues'}, {`disarming errors'}, and {`radio controller issues'}, which are listed in Table \ref{tab:fault_types} as `forum only' issues. 

As with all forum discussions, interpretations and solutions are provided by both experts and non-experts, and represent their opinions about root causes \cite{experts}. As these opinions are not necessarily correct, we therefore read each discussion carefully including the disagreements between contributors, and focused on the final solution and expertise level of the contributor to identify the problem analysis deemed to be the correct one \cite{DBLP:journals/corr/abs-2101-02830}. For discussions that provided both a clear solution and an associated flight log, we retrieved, stored, and annotated the log to show the start and end points of the identified anomaly.

%% file: tables/faults.tex
\begin{table}[]
    \centering
     \caption{Commonly discussed failure types showing information sources and the number of issues associated with each fault type in ArduPilot (AP) and Px4 forums. }
    \label{tab:fault_types}
    \small
    \addtolength{\tabcolsep}{-2.6pt}
    \begin{tabular}{ |L{3.4cm} | C{.5cm} C{.5cm} C{.5cm}| C{0.7cm} C{0.7cm} C{0.7cm}|}
   \hline
        \multirow{2}{*}{\bf Fault Type} & \multicolumn{3}{c|}{\bf Source} & \multicolumn{3}{c|}{\bf Forum Issues}\\ 
       &{\bf CP} & {\bf SC} &{\bf FM} &{\bf AP} &{\bf {Px4}} &{\bf Total}  \\ \hline
         

        Mechanical Failure &  $\CIRCLE$&&$\CIRCLE$& 26&17&43  \\ \hline
        
        Excessive Vibration& $\CIRCLE$& &$\CIRCLE$& 18&23&41 \\ \hline
        
        Compass Interference&  $\CIRCLE$& &$\CIRCLE$ & 17&7&24\\ \hline
        
        Power Issues&  $\CIRCLE$& $\CIRCLE$ &$\CIRCLE$ &13&3&16\\ \hline
        
        GPS Failures&  $\CIRCLE$& $\CIRCLE$ &$\CIRCLE$&6&7&13  \\ \hline

        Failsafe Errors&  $\CIRCLE$&$\CIRCLE$ &  $\CIRCLE$&  8&5&13 \\ \hline       
        
        Barometer Failures& &$\CIRCLE$ &$\CIRCLE$& 3&3&6  \\ \hline
        
        Compass Failures& &$\CIRCLE$ &$\CIRCLE$ &7&5&12 \\ \hline
        
        Inertial Nav. Sys (INS)& &$\CIRCLE$ & $\CIRCLE$ &3&4&7 \\ \hline
        
        Tuning errors& & &$\CIRCLE$ & 10&6&16\\ \hline
        
        EKF switching& & &$\CIRCLE$& 4&0&4 \\ \hline

        Voltage sensor calibration& & &$\CIRCLE$& 3&2&5  \\ \hline
        
        Logging issue& & &$\CIRCLE$& 3&1&4\\ \hline

        Disarming Errors& & &$\CIRCLE$& 0&4&4 \\ \hline  
        
        Radio Controller&&$\CIRCLE$&$\CIRCLE$&2&1&3 \\ \hline
    \end{tabular}
    \\[1ex]
    {\bf Source legend: } CP=Common Diagnosis problems in AP documentation, SC=Safety Checks in AP documentation, FM={AP and Px4 Discussion forums}.

\end{table}

%% file: sec_common_failures.tex
\subsection{Specifying Requirements for Anomaly Detection Models}
\label{sec:common_failures}

In order to evaluate our approach and create an initial set of DLMs, we analyzed five types of failures that were discussed particularly frequently in the discussion forums. These were mechanical problems, vibration problems, electrical interference on the compass, power problems, and GPS connectivity. For each of these we carefully reviewed the documentation and forum discussions to identify the following: {(1) Sensor attributes that serve as indicators for the failure,} (2) Rules for acceptable upper or lower bounds on individual attribute values, or on interactions between attributes (e.g., deltas or combinations) that are indicative of these failures,  and (3) Temporal patterns claimed as indicators of emergent problems.\vspace{4pt}

We then documented our findings using a template that included: failure type, placeholder requirements, and links to alternate causes. Each individual requirement was structured using the EARS event-driven template \cite{EARS} as follows: \vspace{4pt}\newline \noindent \mybox{lightgray}{{\bf When [event] then [named anomaly] shall be detected.}}

The requirements were treated as placeholders because they included qualitative adjectives (highlighted in yellow) such as {\it sufficiently} or {\it suddenly} and/or used unsubstantiated threshold values (highlighted in green) such as {\it more than 100 times} or {\it exceeds 60ms$^2$}. While, we expect the Deep learning model to learn its own patterns and threshold values from the data, these descriptions are useful  for two purposes. First, they can be used later in the process to generate explanations of detected anomalies in terms that are meaningful to humans, and second, they help guide the manual process of annotating flight logs with anomaly tags, thereby constructing a training and evaluation dataset.  

\subsection{An Illustrated Example: Mechanical Problems}
To illustrate our approach we provide a detailed explanation of how requirements were identified and specified for one particular anomaly type related to detecting mechanical problems.

\subsubsection{Analysing forum discussions for posts related to mechanical problems} 
We retrieved 26 different issues discussing mechanical problems from the Ardupilot forum. These primarily referred to motor or ESC failure, broken or loose propellers, imbalance in weight distribution, overly heavy payloads, and improper calibration. Experts agreed that mechanical errors generally cause a sudden divergence of actual attitudes (i.e., roll, pitch, and yaw) from the desired ones represented by the attributes ATT.Roll, ATT.Pitch and ATT.Yaw, and ATT.DesRoll, ATT.DesPitch and ATT.DesYaw. They also discussed that imbalance in RPM values sent to each of the motors, could be indicative of mechanical imbalance caused by uneven weight distribution, a twisted motor mount, or erroneous motor positioning. 

\subsubsection{Specifying requirements for detecting mechanical anomalies} {The expert's comments provided the information we needed to (a) define the symptoms of mechanical problems (i.e., divergence of actual and desired roll, pitch, and yaw), (b) identify associated time-series data streams (e.g., ATT.Roll, ATT.DesRoll), and (c) provide the basis for explaining anomalies related to mechanical problems. Furthermore, in the cases where logs were provided to support the discussion, we retrieved and annotated the log to mark anomalies discussed by the experts, and then saved it for training and testing purposes.}
\input{tables/flight-problems-1}

{As an example, consider the information collected for mechanical failures in Table \ref{tab:reqs_mechanical}. The first requirement states that ``When the actual roll, pitch, or yaw deviates suddenly and sufficiently from desired roll, pitch, or yaw, then an attitude divergence anomaly shall be detected.'' This tells us that we need to build a deep learning model capable of detecting this type of divergence. As the DLM will learn its own thresholds for ``suddenly'' and ``sufficiently'', we accept fuzzy (soft descriptions) in the requirements definition. Our template also includes a section for symptoms, which are cross-referenced with the requirements.} The \emph{symptoms} discussed in the forum highlight the attributes of interest, and allow us to specify a derived requirement (aka design specification).  For example, in the case of Mechanical Failures, we document the fact that we can detect the sudden divergence of actual roll, pitch, and yaw, from desired roll, pitch, and yaw by building a deep learning model based on the time-series data associated with ATT.Roll vs ATT.DesRoll; ATT.Pitch vs ATT.DesPitch; and ATT.Yaw vs. ATT.DesYaw.

\input{tables/flight-problems-2}

\subsubsection{Non-functional requirements for mechanical anomaly detectors} 
{While the primary focus of \RM is on functional, rather than non-functional requirements (NFRs), the \RM template could also be extended to include NFRs. These NFRs may differ significantly according to how and where the DLM will ultimately be deployed.  For example, the DLMs that we developed for the sUAS domain have two distinct purposes.  First, they could be deployed onboard each sUAS to support inflight, real-time analytics and to detect emergent failures and enact mitigations \cite{DBLP:conf/euromicro/VierhauserCBKRG18,DBLP:journals/tse/VierhauserBWXCH21}. Second, they could be used after a failed flight to analyze flight logs to assess the root cause of any failures. 

Each scenario introduces different tradeoffs between qualities such as runtime performance, accuracy, and explainability.  In the onboard scenario, the DLM must support {\it fast} and {\it accurate} analysis.  Two example NFRs might be specified as ``Alerts for anomalous attitude divergence shall be detected and raised within 3 seconds of their occurrence'' and ``Alerts shall be raised at recall and precision levels of 90\%''. In contrast, when used to support post-flight analysis, fast analysis may no longer be a priority, and the DLM might favor recall over precision and require clear explanations of the anomalies to draw attention to potential causes of the flight failure.  In this case, two example NFRs might state that ``Potential anomalies shall be retrieved at recall levels of 99\%'' and ``All candidate anomalies shall provide explanations that are comprehensible to 90\% of RPICs in a randomly recruited user group''. 

Well-established techniques can be used for analyzing quality tradeoffs and specifying NFRs  \cite{robertson,DBLP:journals/ansoft/BoehmEPSKM98,6365165}.}
 The different NFRs for these two use-cases suggests the need to train an individual DLM for each scenario. However, in the remainder of this paper, we focus on developing DLMs that achieve high degrees of accuracy. 

\subsection{Additional Examples of Failure Types}
In order to illustrate the breadth and depth of requirements-related information available in discussion forums, we show four additional examples of RESAM templates in Table \ref{tab:morefailures}. The additional failures include vibration problems, compass interference, GPS glitches, and power issues, all of which represent commonly occurring flight failures, with clear time-series attributes that serve as indicators.   We summarize the expert advice as follows:

\begin{itemize}[leftmargin=*]
 \setlength\itemsep{.2em}
\item {\bf Vibration problems:} 
 Experts explained that vibrations introduce errors in the way accelerometer values are computed, resulting in deviations between GPS readings and the estimated position (computed via GPS readings and barometer) which potentially results in fly-away behavior under certain flight modes  (See Table~ \ref{tab:reqs_vibration}).
 
 \item {\bf GPS glitches} have multiple causes. Experts stated that they  can generally be detected by checking the number of satellites and the horizontal dilution of precision information (See Table~\ref{tab:reqs_gps}).

\item {\bf Compass problems:} Experts focused on electrical interference from onboard components, and noted that the most common case occurs when the throttle is raised and the motors draw high current from the power source. The resulting current can impact magnetometer readings and cause the compass to provide wrong headings, resulting in erroneous yaw estimates and causing the UAV to deviate from its intended flight route (See Table~ \ref{tab:reqs_compass}).

\item {\bf Power brownouts:} Experts again identified multiple causes, with erratic swings in altitude serving as a predictive indicator when the sUAS is temporarily deprived of power (See Table~\ref{tab:reqs_power}).\vspace{0pt}
\end{itemize}

These findings clearly show that leveraging both documentation and forum discussions led to a richer understanding of the attributes that serve as indicators of UAV failures. 

%% file: tables/flight-problems-1.tex
\sethlcolor{ylw}

\begin{table}[h!]

\caption{Placeholder requirements and their supporting time-series data attributes for detecting various forms of anomalies. Qualitative descriptions will ultimately be learned more precisely by the deep learning models. Alternate causes and symptoms serve as placeholders of additional requirements and models. Legend: Yellow = Qualitative descriptors}
\label{tab:reqs1}
     \centering
     \small
     \addtolength{\tabcolsep}{-3.6pt}
     \begin{tabular}{|L{.5cm}|L{4cm}!{\color{lightgray}\vrule}L{3.8cm}|}
     \hline


\multicolumn{3}{|L{8cm}|}{\bf Mechanical Failures}\\ \hline
R1 &\multicolumn{2}{L{7.8cm}|}{ When the actual roll, pitch, or yaw deviates \hl{suddenly and sufficiently} from desired roll, pitch, or yaw, then an \emph{attitude divergence} anomally shall be detected.}\\ \arrayrulecolor{lightgray}\cline{2-3}
& {Sudden divergence of actual attitude (roll, pitch, yaw) from desired attitude.}&ATT.Roll, ATT.Pitch, ATT.Yaw,ATT.DesRoll, ATT.DesPitch,
ATT.DesYaw \\ \arrayrulecolor{lightgray}\hline 
\arrayrulecolor{black}

 R2 &\multicolumn{2}{L{7.8cm}|}{  When an \hl{imbalance} is detected in RPM sent to the motors then an \emph{RPM imbalance} anomaly shall be detected.}\\ \cline{2-3}
  &Imbalanced RPM sent to motors&RCOUT\\ \hline
\multicolumn{3}{|l|}{\bf Alternate Causes of Symptoms}\\ \hline
 1&\multicolumn{2}{|l|}{ Severe wind}\\
 2&\multicolumn{2}{|l|}{ Twitch maneuvers of the UAV (e.g., for collision avoidance).}\\ \hline
     \end{tabular}
      \label{tab:reqs_mechanical}

\sethlcolor{ylw}
\vspace{-8pt}
 \end{table}

%% file: tables/flight-problems-2.tex
\sethlcolor{ylw}
\begin{table*}[t]

\caption{Additional Fuzzy Requirements and their supporting time-series data attributes for detecting various forms of anomalies. Legend: Yellow = Qualitative descriptors, Green = candidate rules.}
\label{tab:morefailures}

\begin{subtable}[t]{0.95\columnwidth}
	\sethlcolor{grn}
     \centering
     \small
     \addtolength{\tabcolsep}{-3.6pt} 
     \caption{Vibration Problems}
      \label{tab:reqs_vibration}\begin{tabular}{|L{.5cm}|L{3.7cm}!{\color{lightgray}\vrule}L{3.5cm}|}
     \hline

R1 &\multicolumn{2}{L{7.8cm}|}{ When the deviation between GPS readings and estimated position exceeds \hl{$30 ms^{-2}$} on any axis (X,Y,Z), then a \emph{geolocation} anomaly shall be detected.}\\ \arrayrulecolor{lightgray}\cline{2-3} 
 & {Standard deviation of accelerometer measurements across three axes.}&VIBE.VibeX, VIBE.VibeY and VIBE.VibeZ  \\ \arrayrulecolor{black}\hline

 R2 &\multicolumn{2}{L{7.8cm}|}{  When the accelerometer reaches its \sethlcolor{grn} maximum limit \hl{more then 100 times} during a mission with \hl{increasing frequency}, then an `overworked accelerometer' error shall be detected.}\\ \arrayrulecolor{lightgray}\cline{2-3}
 &Acceleratometer frequently reaching maximum limit&VIBE.Clip0, VIBE.Clip1 and VIBE.Clip2\\ \arrayrulecolor{black}\hline
 
\multicolumn{3}{|L{8.6cm}|}{\bf Alternate Causes of Symptoms}\\ \hline
 1&\multicolumn{2}{|l|}{ Aging and/or underpowered battery}\\
 2&\multicolumn{2}{|l|}{ Excessive wind.}\\ 
 3&\multicolumn{2}{|l|}{Tuning error (e.g., {\scriptsize{MOT\_THST\_HOVER}} not set correctly)}\\ \hline

     \end{tabular}

      \vspace{8pt}
\end{subtable}
\begin{subtable}[t]{0.95\columnwidth}
	\sethlcolor{grn}
     \centering
           \caption{GPS Glitch}
      \label{tab:reqs_gps}
     \small
     \addtolength{\tabcolsep}{-3.6pt}  
 \begin{tabular}{|L{.5cm}|L{4cm}!{\color{lightgray}\vrule}L{3.2cm}|}
   \hline
   R1 &\multicolumn{2}{L{7.8cm}|}{When GPS.HDop values \sethlcolor{grn}\hl{exceed 2} a `GPS Glitch with Loss of horizontal precision' error shall be raised.}\\ \arrayrulecolor{lightgray}\cline{2-3}
   \arrayrulecolor{black}
   & {Number of satellites }&NSats \\ 
   &Loss of GPS precision.&GPS.HDop\\ \hline
   R2 &\multicolumn{2}{L{7.8cm}|}{When GPS.HDop values \sethlcolor{grn}\hl{exceed 2} and \sethlcolor{ylw}\hl{sudden and sharp} course corrections are detected, then a `GPS Geolocation Failure' shall be raised.}\\ \arrayrulecolor{lightgray}\cline{2-3}
  \arrayrulecolor{black}
  & Sharp flight route divergence&GPS Coordinates \\  & Number of satellites &NSats \\ 
  &Loss of GPS precision.&GPS.HDop\\ \hline
  \multicolumn{3}{|l|}{\bf Alternate Causes of Symptoms}\\ \hline
  1&\multicolumn{2}{|l|}{ Loss of satellite lock}\\
  2&\multicolumn{2}{|L{7cm}|}{Incorrect positioning of components on UAV causing interference}\\ \hline
  \end{tabular}
 \end{subtable}    
\begin{subtable}[t]{.95\columnwidth}
     \centering
     \caption{Compass Interference}
\label{tab:reqs_compass}
     \small
     \addtolength{\tabcolsep}{-3.6pt}
    \begin{tabular}{|L{.5cm}|L{3.7cm}!{\color{lightgray}\vrule}L{3.7cm}|}
     \hline

\hline
R1 &\multicolumn{2}{L{7.8cm}|}{When the a correlation \hl{above 30\%} between the throttle and magenetometers occurs, a \emph{compass interference} anomaly shall be detected.}\\ \arrayrulecolor{lightgray}\cline{2-3}
\arrayrulecolor{black}
& {Increased throttle interferes with compass (detected by correlation between magnetometer readings and throttle) }&l2-norm (i.e.,$\sqrt{SumOfMags}$, where SumOfMags=$MagX^2 + MagY^2 + MagZ^2$) \newline Throttle (CTUN.ThO) \\  \hline
\multicolumn{3}{|l|}{No known alternate causes of symptoms}\\ \hline
\end{tabular}
\end{subtable} 
\hspace{\fill}
\begin{subtable}[t]{.95\columnwidth}
\sethlcolor{grn}
         \caption{Power Issues}
      \label{tab:reqs_power}
      \centering
     \small
    \addtolength{\tabcolsep}{-3.6pt}
\begin{tabular}{|L{.5cm}|L{4cm}!{\color{lightgray}\vrule}L{3.0cm}|}
\hline

R1 &\multicolumn{2}{L{7.8cm}|}{When \sethlcolor{ylw}\hl{increases} in throttle are \sethlcolor{ylw}\hl{correlated with} battery drain then a `throttle causing excessive battery drain' error is detected. }\\ \arrayrulecolor{lightgray}\cline{2-3}
\arrayrulecolor{black}
&Increases in throttle correlated with battery drain&CTUN.Tho, BAT.Volt\\ \hline

 R2 &\multicolumn{2}{L{7.8cm}|}{When erratic swings in altitude are detected then an `altitude fluctuation' error is detected.}\\ \arrayrulecolor{lightgray}\cline{2-3}
 \arrayrulecolor{black}
  & {Erratic swings in altitude}&BARO.Alt, CTUN.Alt, GPS.Alt  \\ \hline
\multicolumn{3}{|l|}{\bf Alternate Causes of Symptoms}\\ \hline
 1&\multicolumn{2}{|l|}{ Excessive wind.}\\
 2&\multicolumn{2}{|l|}{ Onboard software or sensor drain}\\ \hline
     \end{tabular}
\end{subtable}     
 \end{table*}

%% file: sec_techniques.tex
\section{Validating DLM Requirements}
\label{sec:techniques}
Requirements are typically validated for characteristics such as completeness, consistency, validity, ambiguity, and correctness through activities that include requirements analysis, testing, prototyping, and reviews;  however, as the work in this paper represents a preliminary proof-of-concept, we focus only on evaluating whether individual anomaly detectors, designed and constructed to satisfy requirements for detecting a specific type of anomaly, are able to accomplish that goal successfully. 
Our first question therefore evaluates whether the requirements and supporting information such as symptoms and flight logs, provided sufficient guidance and support for building effective DLMs.\vspace{4pt}

\noindent{\bf (RQ1)~{Is the information retrieved from the forums, including the problem definitions, discussions, conclusions, graphs, and flight logs, sufficient for specifying requirements and identifying time-series attributes needed to design and deploy DLMs capable of effectively detecting the targeted anomalies?}} \vspace{4pt}

Next, given the additional time and effort needed to mine requirements-related information from the discussion forums, we posed a second question about the performance of these models versus the models derived solely from the threshold-driven information in the online documentation. Our research question addresses whether the additional effort in eliciting requirements and their supporting data from the forums was worthwhile. The question is stated as follows:\vspace{4pt}

\noindent{\bf (RQ2)~Do the DLM's constructed using data collected from both documentation and forums outperform the rule-based anomaly detectors based on guidance related to threshold values described in the documentation?  }\vspace{-4pt}\newline

We addressed RQ1 and RQ2 by creating two different anomaly detector models for each of three failure types - a subset of mechanical issues exhibiting attitude problems, excessive vibration, and compass interference.  We first trained a Long Short Term Memory (LSTM) autoencoder \cite{LSTMAuto, LSTM}, which  is a variant of the Recurrent Neural Network (RNN) and capable of handling long term temporal dependencies \cite{LSTM} as found in flight log data \cite{autoencoder1,autoencoder2, autoencoder3}; and then created a rule-based anomaly detector using threshold values for sensor data as  defined in the Ardupilot documentation.  For RQ3 we provide only a single preliminary example showing the viability of using \RM data to explain anomalous behavior.

\subsection{Data Sets}
For training and validation purposes we collected 670 Ardupilot dataflashlogs from 23 of our own Ardupilot UAVs including Iris 3DRs, America Hexcopter Spreading Wings S900, and Intel Aeros; and then for  testing purposes we used 46 of the dataflash logs collected from the Ardupilot Discussion Forum (cf., Step S7, Fig. \ref{fig:requirements_model}). All logs were converted to text format using the open source {\it pymavlink} library.  Given the time needed to annotate all 670 training logs, we first eliminated logs that (1) did not include the desired attributes (e.g., VIBE, ATT), (2) had higher degrees of uncertainty expressed in the related forum discussions, (3) had no obvious anomalous behavior matching the targeted attributes, or (4) were smaller than 2 megabytes as larger files tended to provide larger data samples of the anomalous behavior. 350 logs were retained following the filtering process, from which we randomly selected 135 logs for  vibration analysis, 112 logs for attitude analysis, and 38 logs for compass interference. We manually annotated these logs with start and end points of anomalies  associated with attitude, vibration, and compass interference. Table \ref{tab:annotated} summarizes our annotation.

\input{tables/annotated_logs}

\subsection{Constructing LSTM Models }
We trained, validated, and tested each LSTM autoencoder model using Ardupilot flight-logs using the multi-variate time-series data that matched the symptom attributes identified within the \RMx' template. {Autoencoder-based anomaly detection includes several steps.  First (1), we feed the autoencoder non-anomalous data (training data) so that it learns to recognize normal behavior.  Autoencoders take time-series data as input, encode that data, and then learn to decode it back to its original form. They learn to perform this task on non-anomalous data but typically underperform when fed data containing anomalies.  Our second step therefore involves (2) running the model on the training data to calculate the reconstruction error for normal data, and then (3) setting the maximum reconstruction error from the training data as a threshold to distinguish between anomalous and non-anomalous data.  Finally, we (4) apply the model on unseen data with potential anomalies, and detect anomalies when a reconstruction error is returned higher than the established threshold.  Therefore, a higher threshold leads to higher precision but lower recall.}  

For each of the three failure types we extracted the attributes of interest (e.g., VIBE, ATT) from non-anomalous log files in order to build a training set of normal flight behavior in the form of \textit{samples$\times$features}. Through initial trial and error, we selected a sliding window to iterate over the sequences of data in order to create the third dimension - '\textit{timesteps}' required by LSTM, and concatenated data samples from all the non-anomalous log files to create a dataset of shape \textit{samples$\times$timesteps$\times$features}. Then we randomly split the dataset into 80\% for training and 20\% for validation. \vspace{4pt}\newline 

\begin{figure}
     \centering
   \includegraphics[width=1\columnwidth]{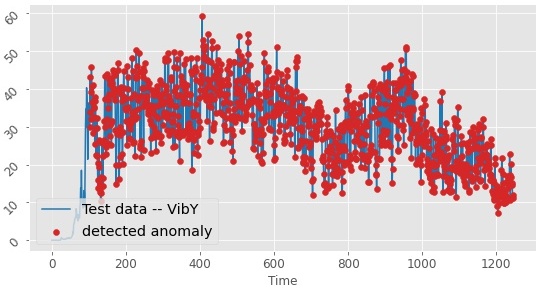}
    \vspace{-8pt}
    \caption{\centering Detection of anomalous vibration by LSTM model}
    \label{fig:res1}
\end{figure}

\noindent{\bf $\bullet$~Vibration:} We used 68 non-anomalous Ardupilot flight logs and extracted the VIBE.VibeX, VIBE.VibeY, VIBE.VibeZ attributes. We experimentally chose a sliding window of 200 and created a dataset of the shape \textit{samples $\times$ 200 $\times$ 3} for each of the log files, and concatenated them to produce a dataset of the shape {255059 $\times$ 200 $\times$ 3}. We applied min-max normalization to make the convergence faster, and used a stacked reconstruction LSTM architecture. The encoder had 2 hidden LSTM layers with 100 and 16 units respectively to create an encoded representation of length 16. A $RepeatVector$ layer was implemented as a bridge between the encoder and the decoder which repeats the encoded representation $timesteps$ (200 in this case) number of times. The decoder served as a mirror to the encoder, and the $TimeDistributed$ layer ensured that the number of output features matched the number of inputs. The $tanh$ activation function was used to help avoid the vanishing gradient problem \cite{tanh}. We used mean absolute error (MAE) loss function and the adaptive moment estimation (Adam) optimizer (learning rate=0.001) was utilized to minimize it. We applied L2 regularization (factor=0.001) to avoid over-fitting, and the model was trained using a batch size of 200 data samples. Visualized results for executing the vibration detection model are shown in Fig.~\ref{fig:res1}, for a flight log with clear vibration problems.

\noindent{\bf $\bullet$~Attitude Anomalies: }
We extracted ATT.DesRoll, ATT.DesPitch, ATT.DesYaw, ATT.Roll, ATT.Pitch, ATT.Yaw from 76 non-anomalous flight logs for training and validation purposes. As we were interested in divergence between values, we subtracted actual values from the desired ones (e.g., ATT.DesRoll - ATT.Roll) for roll, pitch and yaw producing a dataset of the shape \textit{samples $\times$ features} from each of the log files. Using a sliding window of size 200 produced a dataset of shape  {322777 $\times$ 200 $\times$ 3}. 
The Attitude LSTM autoencoder had only one LSTM hidden layer with 64 units on each of the encoder and decoder sides and the \textit{RepeatVector} in the middle. The length of the encoded representation is 64. We used the $tanh$ activation function, MAE loss function and Adam optimizer (learning rate of 0.001). L1 regularization (factor=0.00015) was applied. We fed the data to the model with a batch size of 200.
\vspace{4pt}\newline
\noindent{\bf $\bullet$~Compass Interference: }
 We extracted CTUN.ThO, MAG.MagX, MAG.MagY, MAG.MagZ, MAG2.MagX, MAG2.MagY, MAG2.MagZ from 38 non-anomalous log files. We then calculated the \textit{l2}-norm of the magnetic fields for MAG and MAG2, respectively. Based on initial experimentation, we used a sliding window of size 50 to produce a training dataset of shape {159843 $\times$ 50 $\times$ 3}. The LSTM-based autoencoder has both an encoder and a decoder structure. The encoder has an LSTM layer which encodes the input sequence with 3 features per $timestep$ into 8 hidden features per $timestep$. We take the last hidden representation learned from the input sequence as the encoding. The decoder has an LSTM layer that first converts the dimension of 8 to a dimension of 16 and finally, a fully-connected layer that maps from the hidden space to the original space for reconstruction. We used the MSE loss and an Adam optimizer (learning rate of 0.001) to train the model. We also stacked the input data into a set of mini-batches, where the batch size is 100.

\subsection{Constructing Rule-Based Models}
\label{rule-def}
Our rule-based models use a simple thresholding technique to flag anomalies when the data exceed a predefined threshold. For vibration and compass interference, we use suggested thresholds provided by the Ardupilot documentation. For interference, the threshold value was established when the correlation between CTUN.Tho and the l2-Norm exceeds 30\% and for vibration anomalies, we set a maximum allowable threshold of 30$ms^{-2}$ for each of VibeX, VibeY and VibeZ. As the documentation did not provide thresholds for anomalous attitudes, we tested different thresholds (5$^{\circ}$, 10$^{\circ}$ and 15$^{\circ}$) and selected 10$^{\circ}$ as the one that returned highest accuracy.

\input{tables/results}
\input{figures/graphs}

\subsection{Testing LSTM and Rule-Based Models}
We then applied each model against the test data, which represented previously unseen time-series data from the Ardupilot forum. The vibration test set contained VIBE data for 22 normal and 24 anomalous logs. The attitude test set included ATT  data for 22 normal and 24 anomalous logs. Finally, the compass interference test set included 10 normal and 10 anomalous logs. It was smaller than the others due to fewer available logs.

%% file: tables/annotated_logs.tex

\begin{table}[h]
\centering
\small\addtolength{\tabcolsep}{-1.8pt}
\caption{Annotated Ardupilot Flight Logs with counts of normal and anomalous logs.}
    \label{tab:annotated}

\begin{tabular}{|L{1.8cm}|C{.8cm}C{.8cm}C{.8cm}|C{.8cm}C{.8cm}C{.8cm}|}
\hline

\multirow{3}{*}{\textbf{Type}}
& \multicolumn{6}{c|}{\textbf{Attitude}} 
 \\ 
\cline{2-7}

& \multicolumn{3}{c|}{\textbf{Our own logs}} & \multicolumn{3}{c|}{\textbf{Forum logs}} \\

\cline{2-7}

& \textbf{Total} & \textbf{Norm.} &\textbf{Anom.} & \textbf{Total} & \textbf{Norm.} &\textbf{Anom.}\\ 
\hline
Quadcopters&69&48&21&29&15&14\\ 
\hline
Hexcopters&43&28&15&17&7&10\\ 
\hline

Total&112&76&36&46&22&24\\ \hline
\end{tabular}

\vspace{8pt}
\begin{tabular}{|L{1.8cm}|C{.8cm}C{.8cm}C{.8cm}|C{.8cm}C{.8cm}C{.8cm}|}
\hline

\multirow{3}{*}{\textbf{Type}}
& \multicolumn{6}{c|}{\textbf{Vibration}} 
 \\ 
\cline{2-7}

& \multicolumn{3}{c|}{\textbf{Our own logs}} & \multicolumn{3}{c|}{\textbf{Forum logs}} \\

\cline{2-7}

& \textbf{Total} & \textbf{Norm.} &\textbf{Anom.}& \textbf{Total} & \textbf{Norm.} &\textbf{Anom.} \\ 
\hline
Quadcopters&103&52&51&29&15&14\\ 
\hline
Hexcopters&32&16&16&17&7&10\\ 
\hline

Total&135&68&67&46&22&24\\ \hline
\end{tabular}

\vspace{8pt}
\begin{tabular}{|L{1.8cm}|C{.8cm}C{.8cm}C{.8cm}|C{.8cm}C{.8cm}C{.8cm}|}
\hline

\multirow{3}{*}{\textbf{Type}}
& \multicolumn{6}{c|}{\textbf{Compass Interference}} 
 \\ 
\cline{2-7}

& \multicolumn{3}{c|}{\textbf{Our own logs}} & \multicolumn{3}{c|}{\textbf{Forum logs}} \\
\cline{2-7}

& \textbf{Total} & \textbf{Norm.} &\textbf{Anom.}& \textbf{Total} & \textbf{Norm.} &\textbf{Anom.} \\ 
\hline
Quadcopters&38&30&8&16&10&7\\ 
\hline
Hexcopters&0&0&0&4&0&3\\ 
\hline

Total&38&30&8&20&10&10\\ \hline
\end{tabular}
\end{table}






%% file: tables/results.tex
\begin{table}[]
\centering
\small\addtolength{\tabcolsep}{-2.4pt}
\caption{Area under the ROC curve (AUC) for LSTM autoencoders on different detection attributes}
    \label{tab:auc}
\begin{tabular}{|l|l|l|l|l|l|l|l|}
\hline

\multirow{2}{*}{\textbf{Metrics}}
& \multicolumn{3}{c|}{\textbf{Vibration}}
& \multicolumn{3}{c|}{\textbf{Attitude}}  
& \textbf{Mag Int.} \\ \cline{2-8}

& \textbf{VibeX} &\textbf{VibeY}& \textbf{VibeZ} & \textbf{Roll} & \textbf{Pitch} & \textbf{Yaw} & \textbf{Mag}\\ \hline

LSTM&0.919&0.906&0.958&0.928&0.866&0.817& 0.750\\ \hline

\end{tabular}
\end{table}

\begin{table}[]
\centering
\small\addtolength{\tabcolsep}{-2.4pt}
\caption{\centering F1 measures for the models on the test set using maximum errors on training data as thresholds}
    \label{tab:eval}
\begin{tabular}{|l|l|l|l|l|l|l|l|l|}
\hline

&\multirow{2}{*}{\textbf{Metrics}}
& \multicolumn{3}{c|}{\textbf{Vibration}}
& \multicolumn{3}{c|}{\textbf{Attitude}}  
& \textbf{Int} \\ \cline{3-9}

&& \textbf{VibeX} &\textbf{VibeY}& \textbf{VibeZ} & \textbf{Roll} & \textbf{Pitch} & \textbf{Yaw} & \textbf{Mag} \\ \hline

\multirow{4}{*}{\rotatebox{90}{\textbf{LSTM}}}
&Precision&0.68&0.61&0.86&0.81&0.90&0.83& 0.71\\ \cline{2-9}
&Recall&1.00&1.00&0.95&0.87&0.75&0.63& 1.00\\ \cline{2-9}
&Accuracy&0.87&0.80&0.91&0.89&0.91&0.91& 0.80\\ \cline{2-9}
&F1&0.81&0.76&0.90&0.84&0.82&0.71& 0.83\\ \cline{1-9}

\multirow{4}{*}{\rotatebox{90}{\textbf{Rule}}}

&Precision&0.67&1.00&0.88&0.78&0.83&0.47& 0.53\\ \cline{2-9}
&Recall&0.15&0.35&0.79&0.47&0.42&0.86& 0.80\\ \cline{2-9}
&Accuracy&0.74&0.80&0.87&0.78&0.82&0.80& 0.75\\ \cline{2-9}
&F1&0.25&0.53&0.83&0.58&0.56&0.61& 0.64\\ \cline{1-9}

\end{tabular}

\end{table}

%% file: figures/graphs.tex
\begin{figure}[t]
 
  \begin{subfigure}{0.5\textwidth}
    \centering
    \includegraphics[width=1\columnwidth]{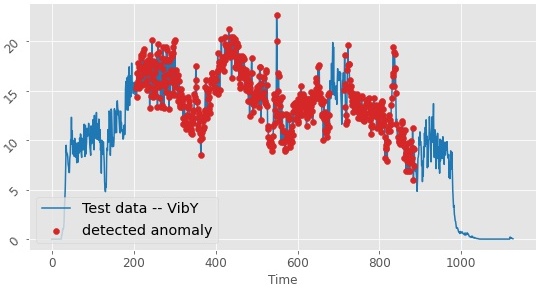}
    \caption{\centering LSTM model successfully detected anomalous vibration below the documented threshold of 30 ms$^{-2}$}
    \label{fig:res3}
  \end{subfigure}
 \\[3ex]
  \begin{subfigure}{0.5\textwidth}
    \centering
    \includegraphics[width=1\columnwidth]{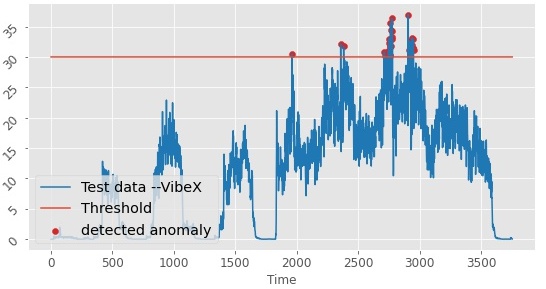}
    \caption{\centering In this example, the rule-based detection method failed to detect vibration anomaly}
    \label{fig:res4}
  \end{subfigure}
    \\[3ex]
  \caption{The LSTM model detected anomalies at higher degrees of accuracy than rule-based methods}
  \label{fig:res}
  \vspace{-8pt}
\end{figure}

%% file: sec_evaluation.tex
\section{{Validation via Analysis of Anomaly Detectors}}
\label{sec:eval}
In this section we first report accuracy results and then discuss their implications with respect to our research questions.

\subsection{Evaluating Results}
We evaluated the efficacy of the LSTM and Rule-based models using the following metrics:

\begin{itemize}[leftmargin=*]
\item \textbf{Area under the Curve:~} For each of the LSTM autoencoders, we measured Area Under the Curve (AUC) of the Receiver Operating Characteristic (ROC) \cite{roc} by plotting the True Positive Rates (TPR) against False Positive Rates (FPR) for a range of autoencoder thresholds. The higher the AUC, the better the model is at differentiating between anomalous and non-anomalous samples \cite{roc2}.

\item \textbf{Precision, Recall, Accuracy and F1 score:~}
For the LSTM autoencoder models we first applied the models to the non-anomalous training data and computed the maximum reconstruction errors for each multi-variate time-series. We used these  reconstruction errors as autoencoder thresholds for detecting anomalies in the test logs. For the rule-based approach, we use thresholds defined in Section \ref{rule-def}. Based on the thresholds established for each model we computed precision, recall, accuracy, and F1-scores with respect to the detected anomalies, where precision is the fraction of detected anomalies that are truly anomalous and recall is the fraction of known anomalies that were correctly detected. F1 score is the harmonic mean of recall and precision, while accuracy is the fraction of correct predictions out of all samples.
\end{itemize}

The AUC results, reported in Table \ref{tab:auc} show that our models performed particularly well for vibration and attitude scoring between ~0.82 and ~0.96 in all cases. The AUC for interference detection was 0.75 which is comparatively lower than vibration and attitude likely because the model needed to learn more sophisticated correlations between two distinct time-series (i.e., throttle and magnetic field). The F1 results in Table  \ref{tab:eval} show that LSTM achieved an average score of ~0.81, compared to that of ~0.57 for the rule-based approach, increasing the F1 score by almost 42\%. 

From Table \ref{tab:eval}, we observe that the LSTM models outperforms the rule-based detection methods for all the attributes. The LSTM models achieved significantly higher F1 score than rule-based detection methods in all the cases.

\subsection{Addressing the Research Questions}
We now address our two requirements-related research questions. We answer RQ1 primarily by discussing the effectiveness of the evaluated DLMs with respect to the stated requirements. Through following our process we found that the forums provided ample information for identifying key anomalies, defining meaningful requirements in the selected template, and identifying time-series data associated with each of the targeted anomalies. While we could have specified actual design definitions, we chose instead to simply list the informal names and formal attribute types of the time-series attributes that would be needed to build a DLM to detect the targeted anomaly. Our experiments served to validate that the requirements provided sufficient information to design, train, validate, and test each DLM, thereby enabling the requirements validation, and ultimately providing support that the RESAM process was effective. 

Our second research question asked whether the additional effort of mining information from the forums was worthwhile, or whether the requirements could have been elicited purely from the formal documentation.  To answer this question we first refer to the types of anomalies depicted in Table \ref{tab:fault_types}, which shows that the documentation discussed six major types of anomalies under `common diagnosis problems', and five (including two overlaps) under `safety checks'. The forums discussed all nine of the faults plus six additional ones.  Furthermore, based on the results from our experiments, the LSTM model consistently outperformed the rule-based detection models by a fair margin in terms of F1 score. We observed three potential flaws in the rule-based detection method. (1) In many cases, actual anomalies occured even though the data remained within the threshold. This is illustrated in Figure \ref{fig:res3} which  shows an example in which anomalous vibration values were below the threshold ($30ms^{-2}$)  even though the vibration was flagged as anomalous by the experts. The rule-based approach failed to detect the anomaly whilst the LSTM model detected it successfully. (2) Rule-based detection often raises false positive results when sudden spikes occur in the data -- for example, quick maneuvers often cause increases in roll, pitch or yaw. LSTM models raised far fewer of these false alarms. Finally, (3) rule-based detection methods detect anomalies only after they exceed predefined thresholds, and are therefore unable to predict emerging anomalies.  For example, in Figure \ref{fig:res4}, the rule-based approach detected an anomaly when the vibration crossed the threshold, whereas the LSTM model was able to forecast the anomaly beginning from the first peak. 

We conclude that the LSTM approach outperformed the rule-based approach. While these findings are clearly influenced by the availability of information on the Ardupilot website and user engagement in the discussion forums, they clearly show that discussion forums can provide useful sources of information for supporting and augmenting the requirements discovery process for anomaly detection.

%% file: sec_discussion.tex
\section{Discussion and Open Challenges}
\label{sec:discussion}
RESAM was created to address the pressing challenge of designing and building onboard analytic solutions for sUAS in the absence of a clear set of requirements. At the start of our project we neither knew which types of flight failure to address, nor which time-series attributes to include in the identified DLMs. Whilst our team had some knowledge of the sUAS domain and runtime monitoring of sUAS systems \cite{SPLC2020,DBLP:journals/tse/VierhauserBWXCH21}, we needed to identify at least a preliminary set of failures to address and then learn how to diagnose those failures based on time-series data attributes available to us.  However, a preliminary analysis of flight control documentation indicated a vast space of over 1400 unique time series data streams exposed by the Ardupilot software.  Without conducting a systematic requirements elicitation process, we would have needed to build an anomaly detector that takes all of these time-series attributes as input, which would require huge amounts of training data and non-trivial resources. This motivated us to explore effective means of supporting the requirements discovery process, ultimately resulting in the RESAM process. 

In retrospect, RESAM gave us the ability to identify and even rank the occurrence of common types of failures, to discover which attributes provide symptomatic clues, and to identify specific data patterns that are indicative of failure. This information enabled us to build an initial suite of DLMs. However, the process we have reported is just a starting point, and the DLMs described here are simply proof-of-concept solutions.  We therefore outline ongoing challenges and opportunities for further research as follows:\vspace{3pt}

\begin{figure*}[t]
    \centering
     \includegraphics[width=.75\textwidth]{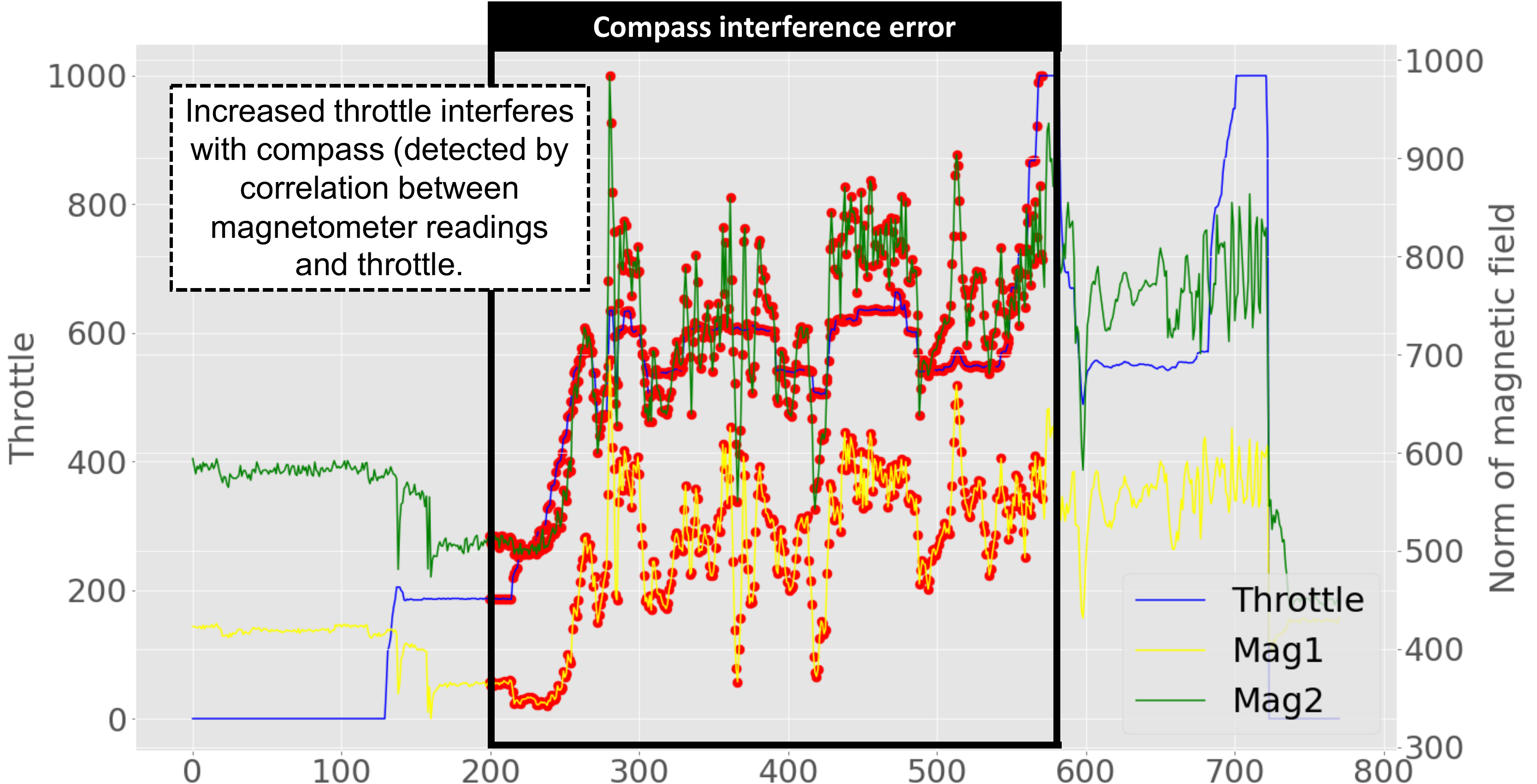}
    \caption{A prototype showing an annotated anomaly, labeled  using explanatory text taken from the \RM template.}
    \label{fig:explainability}
    \vspace{-8pt}
 \end{figure*}

{\begin{itemize}
 \setlength\itemsep{.3em}
\item[{\bf C1:}]{\bf Lack of automation:} Forum mining aspects of RESAM were supported only by basic search features. NLP supported tools are needed that  automatically extract and rank common anomalies, their symptoms, and diagnoses. 
\item[{\bf C2:}]{\bf Complex interactions between symptoms:} In practice a detectable anomaly (e.g., excessive attitude fluctuation) can serve as a symptom of multiple types of failure. Failure diagnosis therefore involves the simultaneous consideration of multiple symptoms, for example, through the use of a decision tree or recommender system \cite{recsys1}. An open challenge, which we are addressing in our current work focuses on automatically mining a far richer set of symptoms and mapping them into a  diagnostic tree. 
\item[{\bf C3:}]{\bf Explainable diagnoses:} Our long-term goal is to explain detected anomalies in ways that are understandable and actionable to RPICs and other stakeholders. The challenge is to mine explanation information from the forums and to integrate this into explanations, as illustrated in Figure \ref{fig:explainability}. In this example an LSTM-detected anomaly is augmented with an explanation extracted from the \RM template.  The algorithms for automating the generation of explanations remains an open challenge, and additional studies are needed to discover how to best explain anomalies to the targeted stakeholders.
\item[{\bf C4:}]{\bf Beyond functional requirements:} As discussed earlier in this paper, NFRs have a strong impact upon the way each DLM is designed and validated, and we need to understand how a DLM will be deployed in order to understand the relevant quality concerns and specify appropriate and measurable NFRs \cite{NFR}. Future work will extend RESAM to support NFRs, potentially through exploring NFR patterns applicable to anomaly detectors, and through developing appropriate tool support.
\item[{\bf C5:}]{\bf Single DLM vs Composition of DLMs:} In this paper we have presented each DLM as an individual entity; however, it is potentially beneficial to combine all of the time-series data, or a subset of critical attributes, into a single DLM and train it to simultaneously detect diverse anomalies. In current work we are comparatively exploring both options. However, from a requirements engineering perspective, the requirements remain the same - to diagnose and explain anomalies as they occur.  
\end{itemize}}

%% file: sec-related.tex
\section{Related Work}
\label{sec:related}
Our research differs from a large body of previous work which has used ML or AI techniques to enhance the general requirements engineering process for specifying, analyzing, and prioritizing requirements in traditional software systems \cite{ml4re1,ml4re2,ml4re3}. In more closely related work, Vogelsang and Borg interviewed four data science experts and identified three key considerations for requirements engineering of ML based systems. These included understanding  qualitative measures specific to ML, addressing new requirement types such as explainability and freedom from discrimination, and generally adapting the RE process for ML \cite{DBLP:conf/re/VogelsangB19}. While their work generally discussed the importance of selecting suitable training data, it did not address the challenges of multi-variate time-series data. Heyn et al., studied AI-intensive Systems Development and identified four RE challenges regarding requirements specification of context, data quality, performance and impact of human factors \cite{wain}. Belani et al., discussed the RE challenges for AI systems and outlined the 'RE4AI' taxonomy for tailoring the AI development process \cite{re4ai}. Finally, Horkoff explored the specification of non-functional requirements (NFR) such as performance, reliability, and maintainability for ML based systems \cite{NFR}. This is relevant to \RM although in this paper we have focused on functionality rather than NFRs. This prior work, which is based on interviews and surveys, discusses the challenges of requirements engineering for ML based systems but does not offer a concrete requirements engineering process for engineering AI-based systems, especially those using time-series data.

We also briefly discuss related work in anomaly detection of time-series data where a large body of prior work exists. Proximity based approaches such as \textit{K}-nearest neighbors \cite{hautamaki2004outlier} and clustering-based methods \cite{cluster, OCSVM} have been proposed as solutions but 
these classifiers do not treat the data as a time-series, rather they treat every sample as a discrete point. Statistical methods \cite{brockwell2009time,https://ARIMA-SARIMA,sarima} capture temporal dependencies, however they are not robust to noise. Deep learning techniques \cite{DL_anomaly, DL_anomaly2}, including  Recurrent Neural Networks (RNN) or Long Short-term Memory (LSTM) structures have been especially popular in time-series anomaly detection because of their ability to capture the patterns of time-series \cite{LSTM-RNN}. LSTM overcomes some of these problems with RNN on long sequences such as the vanishing and exploding gradient problems \cite{tanh,rnndiff}. Using deep learning based autoencoder \cite{autoencoder3,autoencoder1} for anomaly detection has proven to be successful in recent times. LSTM autoencoders have been proven to be very effective in detecting anomalies in time-series data \cite{LSTMAuto}. While out of scope of this paper, we have explored other DL models (e.g., Autoencoder Neural Networks \cite{ann}) to investigate performance vs. accuracy tradeoffs, and will build further on this in our future work.
Further, we are simultaneously exploring a richer multi-variate approach which takes massive amounts of time-series data as input and learns to detect all anomalous behavior but then leverages our \RM to provide human-understandable explanations of the anomalies. This body of work tends to focus upon the actual time-series models without exploring the fundamental requirements engineering processes that are necessary for designing models that meet stakeholders' requirements.